\documentclass[twocolumn,floatfix,tighten]{aastex62}

\received{}
\revised{}
\accepted{}
\submitjournal{ApJ}

\usepackage{graphicx}	
\usepackage{amsmath}	
\usepackage{amssymb}	
\usepackage{bm}		
\usepackage{booktabs}
\usepackage{listings}
\usepackage{color}

\definecolor{dkgreen}{rgb}{0,0.6,0}
\definecolor{gray}{rgb}{0.5,0.5,0.5}
\definecolor{mauve}{rgb}{0.58,0,0.82}

\newcommand{\DAP}{{\tt DAP\,}}	
\newcommand{\pPXF}{{\tt pPXF}}

\newcommand{\samplesize}{$1008 \,$}
\newcommand{\inputsize}{$6507\,$}
\newcommand{\templematchnum}{$5333 \,$}

\let\oldAA\AA
\renewcommand{\AA}{\text{\normalfont\oldAA}}

\usepackage[T1]{fontenc}
\usepackage{ae,aecompl}

\usepackage{newtxtext,newtxmath}

\usepackage{xcolor}
\definecolor{alizarin}{rgb}{0.82, 0.1, 0.26}

\shortauthors{Schaefer et al.}
\shorttitle{SDSS-IV MaNGA: Enriched accretion onto satellites}
\begin{document}

\title{SDSS-IV MaNGA: Evidence for enriched accretion onto satellite galaxies in dense environments}

\author{Adam L. Schaefer}
\affiliation{Department of Astronomy, University of Wisconsin-Madison, 475N. Charter St., Madison, WI 53703, USA}

\author{Christy Tremonti}
\affiliation{Department of Astronomy, University of Wisconsin-Madison, 475N. Charter St., Madison, WI 53703, USA}

\author{Zachary Pace}
\affiliation{Department of Astronomy, University of Wisconsin-Madison, 475N. Charter St., Madison, WI 53703, USA}

\author{Francesco Belfiore}
\affiliation{European Southern Observatory, Karl-Schwarzchild-Str. 2, Garching bei M{\"u}nchen, 85748, Germany}

\author{Maria Argudo-Fernandez}
\affiliation{Instituto de F{\'i}sica, Pontificia Universidad Cat{\'o}lica de Valpara{\'i}so, Casilla 4059, Valpara{\'i}so, Chile}

\author{Matthew A. Bershady}
\affiliation{Department of Astronomy, University of Wisconsin-Madison, 475N. Charter St., Madison, WI 53703, USA}
\affiliation{South African Astronomical Observatory, P.O. Box 9, Observatory 7935, Cape Town, South Africa}

\author{Niv Drory}
\affiliation{McDonald Observatory, The University of Texas at Austin, 1 University Station, Austin, TX 78712, USA}

\author{Amy Jones}
\affiliation{University of Alabama, Department of Physics and Astronomy, Tuscaloosa, AL 35487, USA}

\author{Roberto Maiolino}
\affiliation{University of Cambridge, Cavendish Laboratory, 19 J. J. Thomson Avenue, Cambridge CB3 0HE, United Kingdom}

\author{David Stark}
\affiliation{	Kavli IPMU (WPI), UTIAS, The University of Tokyo, Kashiwa, Chiba 277-8583, Japan}

\author{David Wake}
\affiliation{University of North Carolina at Asheville, Department of Physics,One University Heights, Asheville, NC 28804, USA}

\author{Renbin Yan}
\affiliation{Department of Physics and Astronomy, University of Kentucky, 505 Rose Street, Lexington, KY 40506, USA}

\begin{abstract}
We investigate the environmental dependence of the local gas-phase metallicity in a sample of star-forming galaxies from the MaNGA survey. Satellite galaxies with stellar masses in the range $9<\log(M_{*}/M_{\odot})<10$ are found to be $\sim 0.05 \, \mathrm{dex}$ higher in metallicity than centrals of similar stellar mass. Within the low-mass satellite population, we find that the interstellar medium (ISM) metallicity depends most strongly on the stellar mass of the galaxy that is central to the halo, though there is no obvious difference in the metallicity gradients. At fixed total stellar mass, the satellites of high mass ($M_{*}>10^{10.5} \, \mathrm{M_{\odot}}$) centrals are $\sim 0.1 \, \mathrm{dex}$ more metal rich than satellites of low-mass ($M_{*} < 10^{10} \, \mathrm{M_{\odot}}$) centrals, controlling for local stellar mass surface density and gas fraction. Fitting a gas-regulator model to the spaxel data, we are able to account for variations in the local gas fraction, stellar mass surface density and local escape velocity-dependent outflows. We find that the best explanation for the metallicity differences is the variation in the average metallicity of accreted gas between different environments that depends on the stellar mass of the dominant galaxies in each halo. This is interpreted as evidence for the exchange of enriched gas between galaxies in dense environments that is predicted by recent simulations.

\end{abstract}

\section{Introduction}
\label{sec1}
Star formation in galaxies is maintained through the continued accretion of gas from the intergalactic medium. In turn, the chemical properties of the interstellar medium within galaxies are modulated by the processes of stellar evolution, stellar feedback, and by the abundances of elements within the accreted material. Studies of large samples of galaxies have demonstrated correlations between the gas-phase oxygen abundance in the interstellar medium of galaxies and the total stellar mass \citep{Tremonti2004}, star formation rate \citep{Mannucci2010} and gravitational potential \citep{D'Eugenio2018}. 

These empirical correlations between global galaxy properties and metallicity hint at three processes that can influence the metal content of galaxies. With the buildup of stellar mass through star formation, heavy elements are synthesised in the cores of stars and deposited in the interstellar medium through stellar winds and the terminal phases of the stellar life cycle \citep[see e.g.][]{Maiolino2019}. Assuming that this enriched material mixes efficiently in to the ISM, the metallicity will then depend on the the metal yield from the stars and the amount of gas into which these metals are dispersed. In other words, the metallicity depends on the degree to which the nucleosynthetic products of stellar evolution are diluted in the ISM \citep[e.g.][]{Larson1972,Moran2012,Bothwell2013}. Meanwhile, winds driven by the energy or momentum injected into the ISM through stellar feedback have been shown to remove enriched material from galaxies \citep{Heckman1990}. The magnitude of the effect of metal loss from galaxies due to outflows on setting their metallicity is not known precisely. This is determined by both the mass loading factor, $\lambda$, which quantifies the total mass lost for a given rate of star formation, and also the metal loading factor, which describes the metal content of the outflow. Recent work by \cite{Chisholm2018} found that the the metallicity of outflowing material is independent of the stellar mass of a galaxy, and therefore of the ISM metallicity. However, it is often assumed that the expelled gas has the same metallicity as the ISM \citep{Erb2008,Finlator2008}.

The tradeoff between these different galactic processes has been captured by a number of different attempts to model the chemical evolution of galaxies \citep[e.g.][]{Finlator2008,Peeples2011,Lilly2013}. The \cite{Lilly2013} model makes the simplifying assumption that the evolution of a galaxy`s star formation and metallicity are determined almost entirely by the total gas content. Making simple assumptions about the flow of gas into and out of galaxies, these models are able to reproduce the global mass-metallicity relation, along with the star formation rate - stellar mass relation.

Given the success of analytic models in describing the global properties of galaxies, there has been recent effort to extend the global scaling relations to kpc scales within galaxies \citep[e.g.][]{RosalesOrtega2012,BarreraBallesteros2016,CanoDiaz2016,Medling2018}. These studies have been enabled by the introduction of large-scale spatially-resolved spectroscopic surveys such as the Calar Alto Legacy Integral Field Area Survey \citep[CALIFA][]{Sanchez2012}, the Sydney-AAO Multi-object Integral Field Spectrograph \citep[SAMI][]{Croom2012,Bryant2015} and Mapping Nearby Galaxies at Apache Point Observatory \citep[MaNGA][]{Bundy2015}. \cite{CanoDiaz2016} found that the gas phase metallicity in kpc-sized regions of galaxies is tightly correlated with the stellar mass surface density, $\Sigma_{*}$. However, \cite{BarreraBallesteros2016} showed that this correlation is also dependent on the total stellar mass of the galaxy. That is, at fixed $\Sigma_{*}$ the gas-phase metallicity is correlated with the integrated stellar mass. In direct analogy the argument connecting the global MZR to the increasing depth of the gravitational potential well in more massive galaxies, \cite{BarreraBallesteros2018} showed that the local metallicity also correlates with the local escape velocity. This may explain the connection of the local $\Sigma_{*}$-metallicity relation to the integrated stellar mass.

The existence of local scaling relations are an indicator that some kind of regulatory process is at play, and that processes occurring on local scales may be able to explain the results seen in single-fibre spectroscopic surveys \citep{BarreraBallesteros2016}. Indeed work by \cite{Carton2015} and \cite{BarreraBallesteros2018} found that the \cite{Lilly2013} gas regulator model is able to fit the metallicity given a reasonable estimate of the local gas fraction and mass loading factors.

The ability of gas regulated models to achieve an equilibrium is in part determined by the rate at which gas is accreted and the metallicity of that gas. Historically it was often assumed that the gas fueling star formation is accreted in a chemically pristine state \citep[e.g.][]{Larson1972,Quirk1973,Finlator2008,Mannucci2010}. However, there is mounting evidence that this is not the case \citep{Oppenheimer2010,Rubin2012,Brook2014,Kacprzak2016,Gupta2018}. While work such as that done by \cite{Peng2014} relies on the inference of the properties of gas in the intergalactic medium from modeling, there is a growing body of observations that are able to directly probe the metallicity of gas outside of galaxies by measuring the absorption of background quasar light by extragalactic clouds \citep{Lehner2013,Prochaska2017}. These clouds are observed to be cool and metal rich, and are expected to be accreted onto their host galaxies in the future. Indeed, hydrodynamic simulations \citep{Oppenheimer2010} have shown that below $z \sim 1$ the dominant source of accreted gas onto galaxies is material the was previously ejected.

In addition to internal processes such as star formation and the expulsion of gas by feedback, there are a number of ways in which the gas content of a galaxy could be diminished by external environmental processes. Starvation \citep{Larson1980}, which can occur when a galaxy enters a dense environment and the acquisition of new gas is prevented. In this instance, the existing gas reservoir is consumed by star formation over several Gyr and the metallicity of the ISM increases. Starvation can also be initiated through the heating of gasses in the intergalactic medium by galactic feedback \citep[e.g.][]{Fielding2017}, or by the shock heating of accreted material \citep[e.g.][]{Birnboim2003}. Alternatively, a kinetic interaction between the interstellar medium of a galaxy and the intergalactic medium can result in the ram pressure stripping of gas from a galaxy \citep{Gunn1972}. Occurring on relatively short ($<100 \, \mathrm{Myr}$) timescales, ram pressure stripping is not expected to alter the chemical abundances in a galaxy before it is fully quenched. 

Studies of the environmental effect on galaxy metallicities generally find a small, but significant dependence. For example, \cite{Cooper2008} find that approximately $15 \%$ of the scatter in the mass metallicity relation is attributable to an increase in the metallicity of galaxies in high-density environments. Observationally, there is consensus that the environment has the largest effect on low-mass galaxies \citep{Pasquali2012,Peng2014,Wu2017}. However, the interpretation of this fact is contentious. \cite{Wu2017} attribute the elevated metallicity at fixed stellar mass at greater local galaxy overdensity to a reduction in the gas accretion rate. However, \cite{Peng2014} showed that even when the star formation rates of satellite galaxies in different environments is kept constant, implying no change in the total accretion rate, the metallicity offset is still evident. Their observations and modeling led them to the conclusion that satellite galaxies in dense environments must acquire more enriched gas from their surroundings.

In this paper we make use of the broad range of galaxy environments and stellar masses covered by the MaNGA survey to explore how the local metallicity scaling relations are affected by the environment for satellite and central galaxies. With MaNGA's wide wavelength coverage and spatial resolution we are able to estimate local gas-phase metallicities, gas mass fractions and escape velocities to compare the observations to a model for the gas regulation of the metallicity, and from this modeling infer environmental trends for the metallicity of gas inflows.

In Section \ref{Methods}, we present our data and analysis techniques. In Section \ref{Results} we investigate how galaxy environments change the local metallicity scaling relations, and in Section \ref{Discussion}, we discuss our observations in the context of the \cite{Lilly2013} gas regulator model. Throughout this paper we assume a flat $\Lambda$CDM cosmology, with $H_{0}=70 \mathrm{km \, s^{-1} \, Mpc^{-1}}$, $\Omega_{\Lambda} = 0.7$ and $\Omega_{m} = 0.3$. Unless otherwise stated we assume a \cite{Chabrier2003} stellar initial mass function. We will make use of two oxygen abundance indicators, which each have different absolute abundance scales. The O3N2 \citep{Pettini2004} assumes $12+\log(\mathrm{O/H})_{\odot} = 8.69$ and the N2S2H$\alpha$ indicator \citep{Dopita2016} assumes $12+\log(\mathrm{O/H})_{\odot} = 8.77$.

\section{Methods}\label{Methods}
To investigate the trends of the spatially-resolved metal distribution in galaxies with environment, we make use of the $8$th MaNGA product launch (MPL-8) internal data release, which is similar to the SDSS DR15 \citep{Aguado2018}, but includes data from of \inputsize unique galaxies. In this Section we describe the data, sample selection and methods used to perform our analysis on these data.

\subsection{MaNGA data}
The MaNGA survey is the largest optical integral field spectroscopic survey of galaxies to date. Run on the $2.5 \, \mathrm{m}$ SDSS telescope \citep{Gunn2006} at Apache point observatory, the MaNGA survey aims to observe approximately $10,000$ galaxies. This sample comprises a primary and a secondary subsample that were selected to have approximately flat distributions in the $i$-band absolute magnitude, $M_{i}$. These provide coverage of galaxies out to $1.5$ and $2.5 \, R_{e}$ respectively \citep{Wake2017} and median physical resolutions of $1.37$ and $2.5 \, \mathrm{kpc}$.

Observations of each galaxy are made with one of $17$ hexagonal optical fibre hexabundles, each comprising between $19$ and $127$ $2\arcsec$ optical fibres, subtending between $12\arcsec$ and $32\arcsec$ on the sky. The fibre faces fill the bundle with $56\%$ efficiency, so each target is observed with a three-point dither pattern with $15$-minute exposures per pointing. This pattern of observations is repeated until a median S/N of $20 \, fibre^{-1} \, pixel^{-1}$ is achieved in the $g$-band, which is typically $2-3$ hours in total \citep{Law2015,Yan2016}. Light from each hexabundle is taken from the fibres to the BOSS spectrograph \citep{Smee2013} where it is split by a dichroic at $\sim 6000 \, \mathrm{\AA}$ into red and blue channels, then dispersed at $R \approx 2000$. The resulting spectra are then mapped onto a regular grid with $0.5\arcsec$ square spaxels, with continuous wavelength coverage between $3600 \, \mathrm{\AA}$ and $10300 \, \mathrm{\AA}$. For an in-depth discussion of the MaNGA data reduction pipeline, see \cite{Law2016}.

The reduced data are analysed by the MaNGA Data Analysis Pipeline \citep[\DAP;][]{Belfiore2019,Westfall2019}, which extracts stellar kinematics, measures the strengths of continuum features, and extracts emission line fluxes, equivalent widths and kinematics for each galaxy. For this work, we make use of the emission line fluxes from the \DAP's hybrid binning scheme. In this scheme, the data cubes are Voronoi binned \citep{Cappellari2003} to a S/N of at least $10$ in the continuum. Within each of these bins, \pPXF ~\citep{Cappellari2004} is used to fit an optimal continuum template which is made up of a linear combination of hierarchically-clustered templates from the MILES stellar library \citep{SanchezBlazquez2006,FalconBarroso2011} as well as an $8$th degree multiplicative Legendre polynomial. This optimal continuum template constrains the stellar populations within the Voronoi bin, and is fitted by \pPXF ~in conjunction with a set of Gaussian emission line templates to each individual spaxel in the bin. For a full description of the \DAP ~fitting process, see \cite{Westfall2019}, and for a discussion of the robustness of the emission line measurements, see \cite{Belfiore2019}.

\subsection{Sample Selection}\label{sample_selection}
Galaxies in the MaNGA survey are selected such that the full sample has a roughly flat distribution of stellar masses. However, this parent sample contains galaxies with a wide range of morphologies and star formation rates. Our goal with this work is to make spatially-resolved measurements of the properties of the gas in galaxies as a function of the local environment. The metallicity indicators mentioned in Section \ref{Calibrators} are only calibrated for HII regions, and therefore cannot be applied to a fraction of the spaxels in MaNGA. To make this determination, we compare the $\mathrm{[NII]\lambda 6584 / H\alpha}$ and $\mathrm{[OIII]\lambda 5007 / H\beta}$ emission line ratios on a \cite{Baldwin1981} (BPT) diagram. Only spaxels with emission line ratios that satisfy both the \cite{Kauffmann2003} and \cite{Kewley2001} criteria for excitation of the gas by a young stellar population are included in our analysis. We further exclude spaxels for which the S/N ratio in the emission lines used for the metallicity and determination of star formation are less than $3$. \cite{Belfiore2019} showed that the fluxes of lines above this threshold in MaNGA are relatively robust to systematic effects. From these constraints we calculate the fraction of spaxels for a data cube for which we are able to reliably determine a metallicity. In computing this fraction, we include only spaxels where the $g$-band flux from the data cube is detected at a S/N of $2$ or greater. This condition is imposed so that galaxies that do not fully fill the IFU field of view are not unduly excluded. Galaxies for which the fraction of spaxels with a measurable metallicity is larger than $60\%$ are retained for our analysis.

We make a further cut on the galaxies in our sample based on our ability to robustly measure their metallicity gradients. Galaxies with an elliptical minor to major axis ratio ($b/a$) less than $0.6$, as determined by the NASA-Sloan Atlas \citep[NSA;][]{Blanton2011} Elliptical Petrosian photometry, were excluded to give a sample of face-on galaxies. We made a further restriction on the measured $r$-band effective radius, $R_{e}$. Galaxies with $R_{e}<4\arcsec$ are also rejected. These criteria are motivated by the analysis performed by \cite{Belfiore2017}, who showed that beam-smearing effects are non-negligible for highly inclined systems, or for galaxies that are small relative to the MaNGA point spread function. Similarly, \cite{Pellegrini2019} used a set of realistic simulations to show that light-weighted quantities, such as dust attenuation, are systematically overestimated when observed in highly inclined systems.

\begin{figure}
\includegraphics{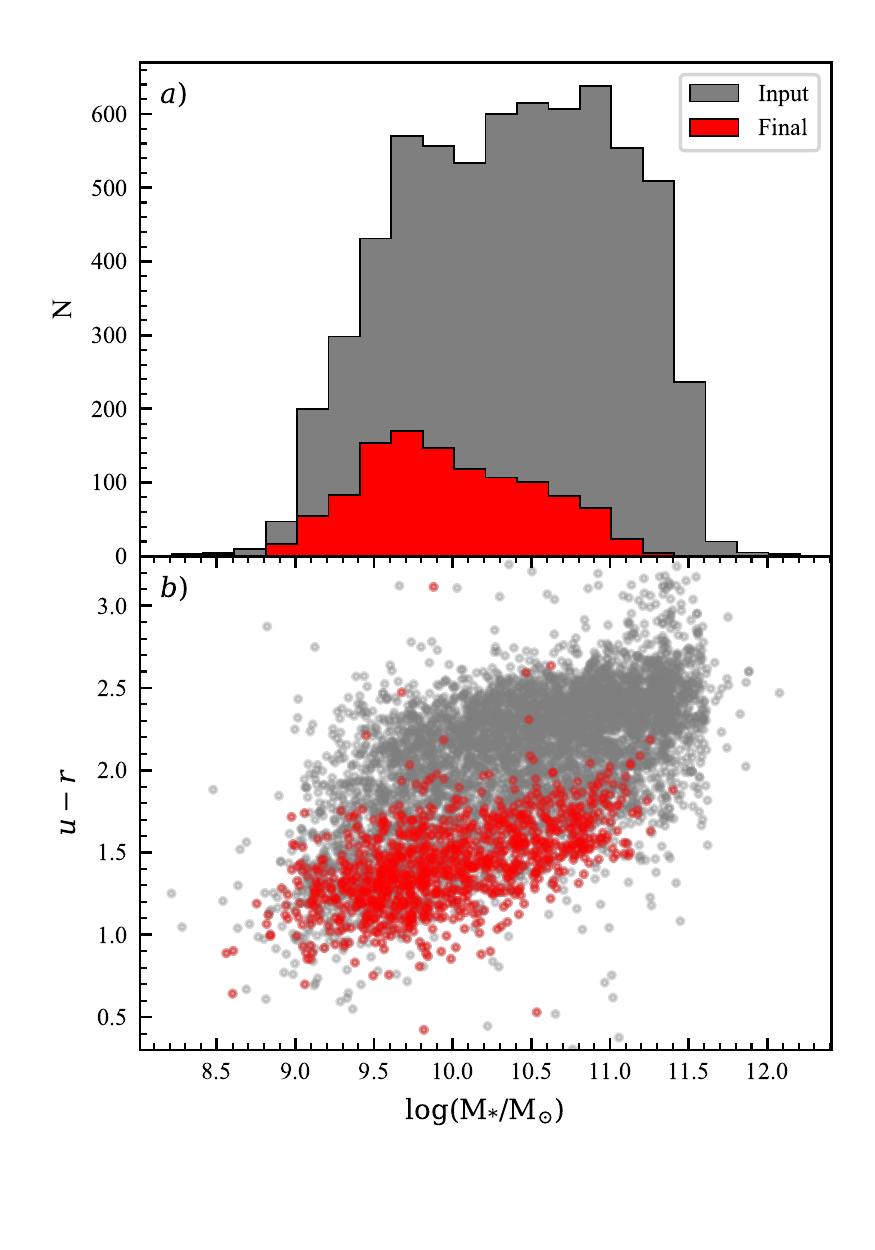}
\caption{In panel $a)$ we show the distribution of stellar mass, $\mathrm{M_{*}}$ for the input sample (\textit{grey}) and for the final sample (\textit{red}). The attrition of sources occurs preferentially at high stellar mass, which is consistent with the rising fraction of passive galaxies. In $b)$ the positions of galaxies in the input sample (\textit{grey}) and final sample (\textit{red}) on the $u-r$ colour-mass diagram is shown. Galaxies that satisfy our selection criteria are predominantly in the blue cloud and forming stars.}\label{M_star_cmd}
\end{figure}

Our final sample consists of \samplesize galaxies, with stellar masses in the range $7.8< \log(M_{*}/M_{\odot})<11.4$. For the majority of our analysis we restrict the stellar masses considered to $9< \log(M_{*}/M_{\odot})<11$, and in this range our sample comprises $967$ galaxies. The distribution of galaxies in our sample the colour-mass plane is shown in Figure \ref{M_star_cmd}.

\subsection{Gas-phase metallicities}\label{Calibrators}
To probe the chemical evolution of the galaxies in our sample, we use the \DAP ~emission line measurements to estimate the gas-phase oxygen abundances in these systems. For convenience we will use the terms `gas-phase oxygen abundance' and `metallicity' interchangeably throughout this work.
The estimation of gas-phase oxygen abundances with optical spectroscopy is often achieved by measuring a set of emission line ratios that vary with the conditions of the gas. The metallicity can be calculated by comparing the measured line ratios to theoretical models for HII regions \citep[e.g.][]{Blanc2015}. Alternatively it can be estimated by using relationships that are empirically calibrated by comparing these line ratios to the spectra of HII regions for which the metallicity has been measured directly using temperature-sensitive emission line ratios, such as $\mathrm{[OIII] \lambda 4363 / [OIII] \lambda 5007}$. While it is generally accepted that the direct method of the oxygen abundance determination is more robust than theoretical modeling or using empirical calibrations, it is generally not possible with datasets such as MaNGA as the $\mathrm{[OIII] \lambda 4363}$ line is typically $\sim 100$ times fainter than the $\mathrm{[OIII] \lambda 5007}$ line.

Many of the empirical calibrations suffer from systematics, being biased either high or low due to variations in the ionisation parameter in the gas, or contamination from diffuse ionised gas or light from an AGN. To account for this fact, we will make use of two different metallicity calibrations. 
\subsubsection{O3N2}
For consistency with \cite{BarreraBallesteros2018}, we use the \cite{Pettini2004} $O3N2$ oxygen abundance diagnostic. This method makes use of the $\mathrm{[OIII] \lambda5007/H\beta}$ and $\mathrm{[NII]\lambda6584 / H\alpha}$ emission line ratios, and was calibrated against a set of $137$ extragalactic HII regions for which a metallicity had been determined either by the direct $T_{e}$ method or by photoionisation modeling of their spectra. Taking $O3N2 = \mathrm{\log([OIII]\lambda 5007 / H\beta) - \log([NII]\lambda6584 / H\alpha)}$, the metallicity of an HII region can be calculated as
\begin{equation}
12+\log(\mathrm{O/H}) = 8.73 - 0.32 \times O3N2,
\end{equation} 
over the range $8.1 < 12+\log(\mathrm{O/H}) < 9.05$. It should be noted that this calibration suffers from some degeneracy with variation in the ionisation parameter within the gas. For this reason it cannot be used in spectra that include significant contamination from diffuse ionised gas, or active galactic nuclei, and may also be biased by variations in $q$ between star-forming regions within a galaxy \citep[see e.g.][]{Poetrodjojo2018}.

\subsubsection{N2S2H$\alpha$}
An alternative method for calculating the metallicity of HII regions based on the relative intensities of the $\mathrm{[NII]\lambda 6584}$, $\mathrm{[SII]\lambda 6717,6731}$ and H$\alpha$ lines was presented by \cite{Dopita2016}. Assuming a simple relationship between N/O and O/H, and modeling theoretical HII regions with a variety of gas pressures and ionisation parameters using the {\tt MAPPINGS 5.0} software, they found
\begin{equation}
\begin{aligned}
12+\mathrm{\log(O/H)} = & 8.77 + \log(\mathrm{[NII]\lambda6584 /[SII]\lambda6717,6731}) \\
	& + 0.264 \log(\mathrm{[NII]\lambda6584 /H\alpha}),
\end{aligned}
\end{equation}
which they showed has very little dependence on the ionisation parameter and is valid over the range $8.0 \lesssim12 + \log(\mathrm{O/H}) < 9.05$. There is some evidence that the relationship between N/O and O/H varies with the total stellar mass of a galaxy \citep{Belfiore2017}, but this should be negligible if analysis is carried out within narrow bins of stellar mass.

Each of the two metallicity calibrations outlined above have different absolute abundance scalings. While it is not possible to directly compare the metallicities of galaxies derived with different calibrations, relative differences between two measurements made with the same indicator are likely to reflect real differences in the chemical composition of the galaxies in question.

\subsection{Gas density}
Following the methodology of \cite{BarreraBallesteros2018} we estimate the local neutral gas surface density from the dust attenuation derived from the observed Balmer line ratios. Under the assumption of a fixed gas-to-dust ratio, \cite{BarreraBallesteros2018} utilised the observation that the gas surface density is related to the V-band attenuation via $\Sigma_{gas} = 30 \times \mathrm{A_{V}} \, \mathrm{pc^{-2}}$. We apply a small correction to this to account for the variation in the dust-to-gas ratio with the gas phase metallicity using the relation given by \cite{Wuyts2011},
\begin{equation}
\Sigma_{gas} = 30 \times A_{V} \times \left( \frac{Z}{Z_{\odot}} \right)^{-1} \, \mathrm{pc^{-2}},
\end{equation}
for $Z<Z_{\odot}$ and is independent of metallicity above $Z_{\odot}$. In this expression we calculate the total extinction in the $V$-band as
\begin{equation}
A_{V} = R_{V} \times \frac{2.5}{k(\lambda_{\mathrm{H\beta}}) - k(\lambda_{\mathrm{H\alpha}})} \log \left(\frac{f(\mathrm{H\alpha}) / f(\mathrm{H}\beta)}{2.86}\right).
\end{equation}

For the \cite{Cardelli1989} extinction law, $k(\lambda_{\mathrm{[OII]}}) = 4.77$, $k(\lambda_{\mathrm{[NII]}}) = 2.53$, $k(\lambda_{\mathrm{H\alpha}}) = 2.54$ and $k(\lambda_{\mathrm{H\beta}}) = 3.61$, and the ratio of total to selective extinction, $R_{V}$, is $3.1$. For our sample (described below), $95\%$ of the spaxels have metallicity above $0.5 Z_{\odot}$, meaning that the majority of gas surface density measurements will differ systematically from those derived by \cite{BarreraBallesteros2018} by a factor of two at most.

\subsection{Stellar mass surface density}
A large fraction of the baryons in the inner parts of galaxies have been locked away in long-lived stars. The stellar content of a galaxy is therefore a valuable diagnostic for its integrated star formation and chemical enrichment history. For this work we take the estimates of stellar mass per spaxel provided by Pace et al. (2019a, 2019b). With this method, individual spectra are fitted using a basis set of six vectors, found using principal component analysis (PCA) of a permissive set of $40,000$ star formation histories. These basis spectra were generated from the C3K library (Conroy, \textit{in prep.}), using a \cite{Kroupa2001} stellar initial mass function. Dust attenuation is taken into account using a two-component model following \cite{Charlot2000}, whereby light from the younger component of the stellar population experiences a different degree of extinction than the light from the older stellar populations. Pace et al. (2019a) showed that this technique provides statistically robust estimates of the stellar mass-to-light ratio across a wide range of signal-to-noise ratios, star formation histories and metallicities, with random uncertainties typically $0.1 \, \mathrm{dex}$ or less for spectra with $\mathrm{S/N}>2$.

We have rescaled the estimated stellar masses from a \cite{Kroupa2001} to a \cite{Chabrier2003} initial mass function by dividing by $1.06$ \citep{Zahid2012}. To calculate stellar mass surface density ($\Sigma_{*}$) in a spaxel, we take these values and divide them by the projected area of a $0\farcs 5$ square spaxel at the systemic redshift the host galaxy. A small correction for inclination of the galaxy is applied by multiplying these surface densities by the elliptical Petrosian minor-to-major axis ratio.

\subsection{Estimating the local escape velocity}
We estimate the local escape velocity from the halo assuming a spherically-symmetric dark matter halo that is described by a \cite[NFW]{Navarro1997} profile using the same procedure as \cite{BarreraBallesteros2018}. This method assumes that the star-forming gas is confined to a thin disk coplanar with the optical disk, and is in a circular orbit around the centre of the galaxy. We extract a rotation curve by taking the maximum and minimum measured line-of-sight velocity within a $30^{\circ}$ wedge along the photometric major axis of the galaxy, similar to \cite{BarreraBallesteros2014}. This rotation curve is corrected for the galaxies inclination to our line of site using the $r$-band elliptical Petrosian major to minor axis ratio. We fit the resulting rotation curve using the \cite{Bohm2004} parametrisation,

\begin{equation}
V(r_{depro}) = V_{max} \frac{r_{depro}}{\left(R_{turn}^{\alpha}  + r_{depro}^{\alpha}    \right)^{1/\alpha}},
\end{equation}
where $V_{max}$ is the maximum velocity of rotation, $r_{depro}$ is the deprojected radius, $R_{turn}$ is the radius at which the rotation curve flattens, and $\alpha$ is a parameter that determines the shape of the rotation curve. This formulation is a special case of the phenomenological model presented by \cite{Courteau1997}. The fitted parameters $V_{max}$ and $R_{turn}$ are then used to derive the local escape velocity using the following formula:
\begin{equation}
 V_{esc}^{2} = 
	\begin{cases}
		V_{esc,in}^{2} + V_{esc,out}^{2} & r_{depro} < R_{turn} \\
		V_{esc,out}^{2} & r_{depro} > R_{turn}
	\end{cases} 
\end{equation}
where
\begin{equation*}
V_{esc,in} = \left(V_{max}/R_{turn}\right)^{2} \left(R_{turn} - r_{depro} \right)^{2} + V_{esc,out}^{2}
\end{equation*}
and
\begin{equation*}
V_{esc,out}^{2} = 2V_{max}^{2} \ln \left( R_{vir}/r_{depro}     \right) + 2V_{max}^2.
\end{equation*}
In this relation, $R_{vir}$ is the virial radius of the galaxy's halo, which we obtain by estimating the galaxy's total halo mass from its stellar mass using the relation derived by \cite{Behroozi2010}. This computation of the local escape velocity assumes the galaxy potential to be spherically symmetric. \cite{BarreraBallesteros2018} tested how a more complicated two-component halo, that includes a contribution to the gravitational field from the baryons in galaxy disk and found that this causes a deviation in the estimated escape velocity of only $\sim 5\%$  from the simpler spherical case.

\subsection{Environment}
\begin{figure}
\includegraphics{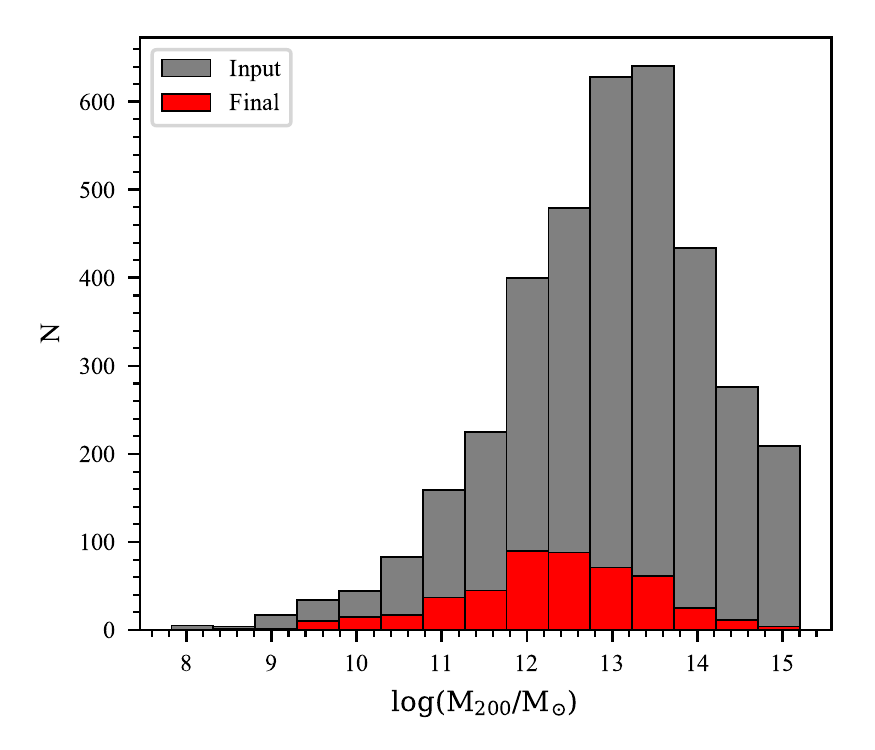}
\caption{The distribution of group halo mass, $M_{200}$ for the input sample (\textit{grey}) and for the final sample (\textit{red}). The loss of sources from the input sample at higher halo mass is more severe than in low mass halos due to the higher fraction of passive galaxies in the most dense environments. }\label{M_200_dist}
\end{figure}

The aim of our current work is to explore how the local environments that galaxies are in today impact their chemical evolution. While there are many different ways of characterising environment, each capable of tracing a variety of different physical processes that can occur during a galaxy's lifetime, we will use the satellite/central classification of our sample as the primary metric for environment. This has been shown to be a good predictor of the star-forming properties of galaxies at fixed stellar mass \citep[e.g.][]{Peng2012}. We make use of the \cite{Tempel2017} catalogue, which uses a friends-of-friends algorithm to provide estimates of group membership, group richness, and dark matter halo mass for galaxies in SDSS DR12 \citep{Eisenstein2011, Alam2015}. Of the \inputsize galaxies in MPL-8, \templematchnum were associated with groups in the \cite{Tempel2017} catalogue, of which $3447$ are identified as the centrals of their halo and $1886$ are satellites. This sample comprises a wide range of group masses, $M_{200}$, which is the mass contained within $R_{200}$ of the group centre, the radius at which the density of an NFW profile drops to $200$ times the average density of the Universe. Groups within this catalogue contain as few as two galaxies and up to $254$ members for the most massive halo. We show the distribution of halo masses for our final sample in Figure \ref{M_200_dist}.

\begin{figure*}
\begin{center}
\includegraphics{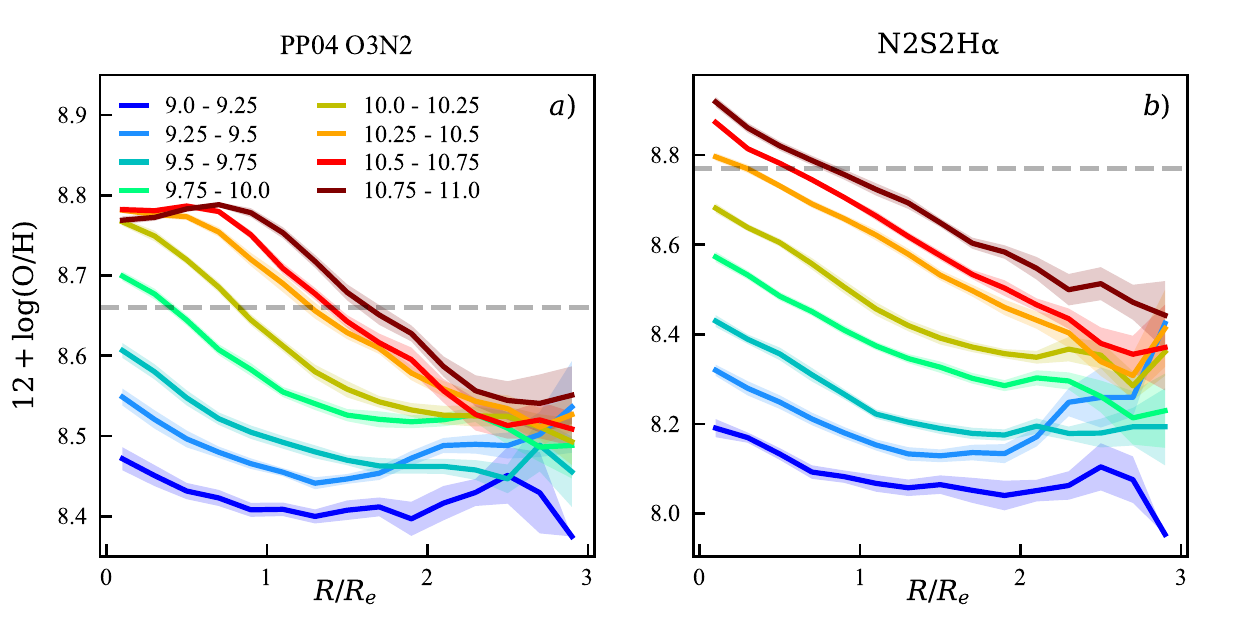}
\end{center}
\caption{The median metallicity radial profile for galaxies in the stellar mass ranges indicated by the legend at the top of panel $a)$. The solid curves represent the median profiles, while the shaded regions of the same colour represent the $1\sigma$ error range on the median. In panel $a)$ we show the metallicity median profiles made using the \cite{Pettini2004} O3N2 indicator and in panel $b)$ we show the results from the \cite{Dopita2016} $\mathrm{N2S2H\alpha}$ indicator. Note that each metallicity indicator has a different abundance scaling on the y-axis, and in each panel we mark the assumed solar abundance with a grey dashed line.}\label{RP_vs_mass}
\end{figure*}

\subsection{Calculating metallicity radial profiles}
To characterise the radial dependence of metallicity in the galaxies in our sample, we construct the radial profile in the following way. In each galaxy, the deprojected distance from the centre of the galaxy has been calculated by the \DAP based on the $r$-band surface brightness distribution in the SDSS imaging, and assuming each galaxy to be a tilted thin disk. We measure the metallicity as a function of radius and take the average value for spaxels in $0.5\arcsec$-wide bins. The metallicity measurements are included only if they are classified as star-forming on the BPT diagram, if they have $\mathrm{S/N} >3$ in all emission lines utilised, and if the $\mathrm{H\alpha}$ equivalent width is greater than $6 \, \mathrm{\AA}$ in emission. This $\mathrm{H}\alpha$ equivalent width criterion is consistent with that chosen by \cite{BarreraBallesteros2018} to minimize contamination of the emission line fluxes from diffuse ionised gas.

To understand the behaviour of an ensemble of galaxies we measure what we will call the `median profile' for the metallicity. For this, we take the median of the individual radial profiles within radial bins that are $0.2 \, R_{e}$ wide. Once the median has been calculated we perform a bootstrap resampling of the galaxies, recalculating the median profile for $1000$ realisations of the sample. At each radius, the uncertainty on the sample median is estimated to be the standard deviation of the bootstrapped median profiles. 

\section{Results}\label{Results}

\subsection{Metallicity profiles}
Following \cite{Belfiore2017} we calculate the median metallicity profiles in narrow bins of stellar mass using two different oxygen abundance indicators. The median profiles for all galaxies in our sample are shown in Figure \ref{RP_vs_mass}. In panel $a)$ of this figure, we show the \cite{Pettini2004} O3N2 oxygen abundance median profiles in $0.5 \, \mathrm{dex}$-wide bins of stellar mass. The profiles shown here are consistent with those shown in Figure 3 of \cite{Belfiore2017}, in particular with the steepening of the metallicity gradient at higher stellar mass. Panel $b)$ shows the profiles for the same galaxies derived using the \cite{Dopita2016} $\mathrm{N2S2H\alpha}$ indicator. This indicator shows qualitatively different results to O3N2 in the centres of galaxies above $\log(M/M_{\odot}) \sim 10.25$. While the $\mathrm{N2S2H\alpha}$ metallicities both continue to rise in the centres of massive galaxies, the O3N2 indicator shows a flattening. The origin of the mismatch in the behaviour of the metallicity profiles in the centres of massive galaxies between different abundance calibrations may be due to the fact that the \cite{Dopita2016} metallicity calibration is strongly tied to the N/O ratio. \cite{Belfiore2017} showed N/O to increase towards the centres of massive galaxies, while O/H does not.

Again, we caution that the interpretation of metallicity radial profiles from datasets with kpc-scale physical resolution such as MaNGA is subject to the flattening of gradients by the observational point spread function \citep{Yuan2013,Mast2014}. \cite{Carton2017} presented a method to account for this effect, however they reported that it was not always robust in the presence of clumpy star formation distributions. Our galaxy size and inclination selection criteria that were outlined in Section \ref{sample_selection} should mitigate the most severe resolution effects \citep{Belfiore2017}, but we note that the most accurate determinations of metallicity gradients require observations with finer spatial resolution. The core conclusions of this work, particularly those based on local scaling relations, will be only minimally affected by this issue.

\subsubsection{Satellites vs. centrals}
\begin{figure*}
\includegraphics{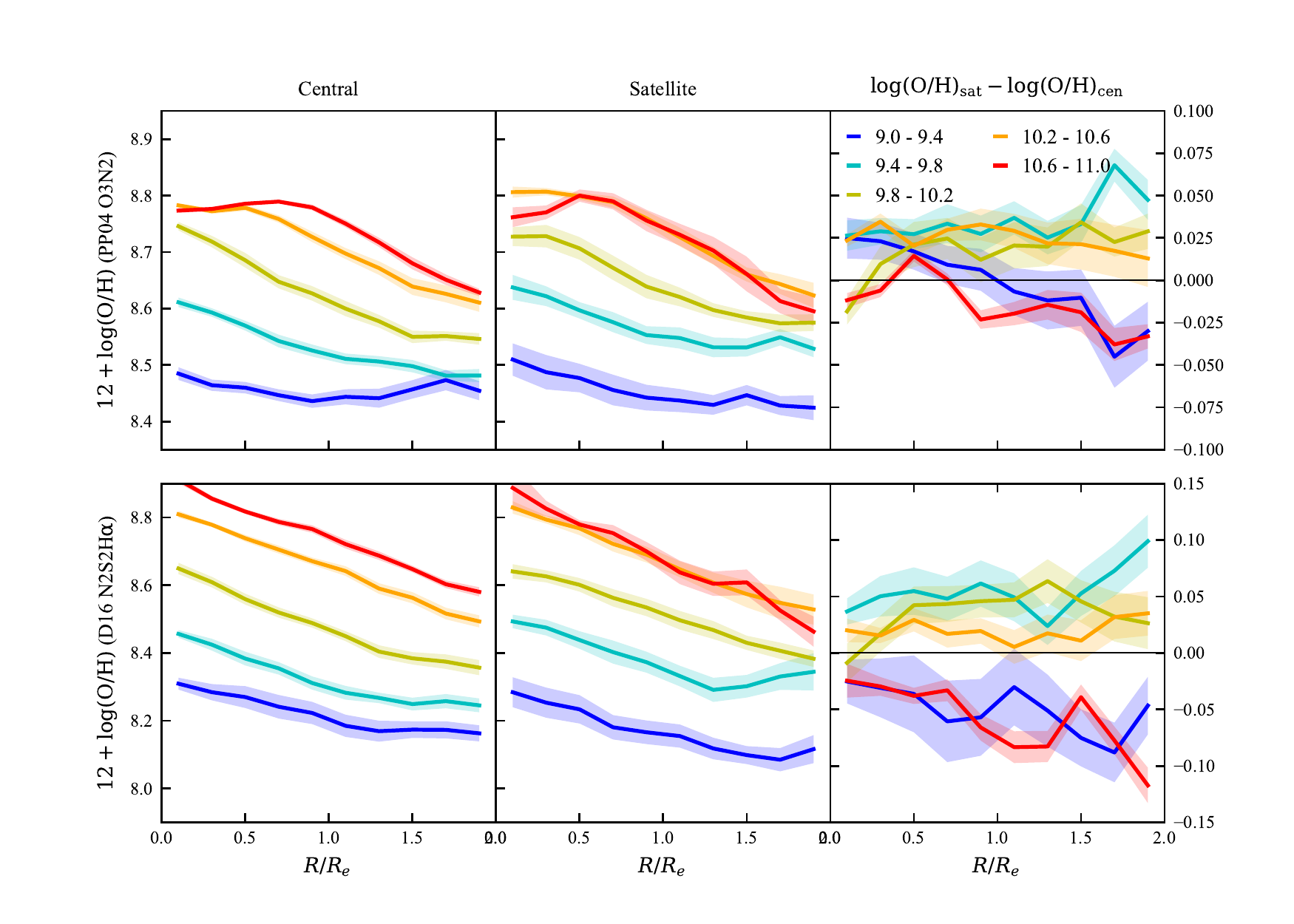}
\caption{The median metallicity radial profiles in bins of stellar mass split into satellites and centrals. In the upper row, we show the profiles for the O3N2 indicator, while in the lower row the results for $\mathrm{N2S2H\alpha}$ are shown. The profiles for central galaxies are shown in the left column, and for the satellite galaxies in the middle column. On the right we show the difference between the satellites and centrals. Satellite galaxies in the range $9.4 < \log(M_{*}/M_{\odot})<10.2$ are systematically more metal rich than centrals of the same mass in both metallicity indicators.}\label{sat_cen_rp_all}
\end{figure*}

To investigate the environmental dependence of the radial distribution of the oxygen abundance, we split our sample into satellites and centrals, then re-calculate the median profiles with each metallicity indicator. We show these radial profiles in Figure \ref{sat_cen_rp_all}. At fixed stellar mass there are minor qualitative and quantitative differences between the radial distributions of the oxygen abundance. At all radii, the absolute differences in the median metallicity at fixed stellar mass are less than $\sim 0.05 \, \mathrm{dex}$ for O3N2, and less than $\sim 0.1 \, \mathrm{dex}$ for $\mathrm{N2S2H\alpha}$. We note that in the highest stellar mass bins, there are very small numbers of star-forming satellites, and their distribution is biased towards the lower stellar masses. For this reason the differences for the most massive galaxies are not robust in this sample. At lower stellar masses, however, the stellar mass distributions of satellites and centrals are similar, the sample sizes are larger and a fairer comparison can be made.

Since each metallicity indicator has different systematics and biases, we only deem a difference to be real if it is reflected in both the O3N2 and $\mathrm{N2S2H\alpha}$ data. For galaxies in the mass range  $9.4 < \log(M_{*}/M_{\odot})<10.2$, there is a systematic offset in the metallicity, with satellite galaxies being more metal rich at all radii sampled, but with no significant difference in the gradient. In the lowest mass range, the O3N2 indicator shows a change in the metallicity gradient, however this is not evident in the $\mathrm{N2S2H\alpha}$ metallicity.

\subsection{Local scaling relations}\label{local_scaling_relations}
In addition to the global scaling relations relating a global metallicity measurement to the integrated stellar mass of a galaxy, there are also local correlations between the stellar mass surface density \citep{Moran2012,RosalesOrtega2012} and the local gas-phase metallicity. These local scaling relations capture the response of the chemical abundance of gas to processes occuring on local $\sim \mathrm{kpc}$ scales. 

\begin{figure*}
\includegraphics{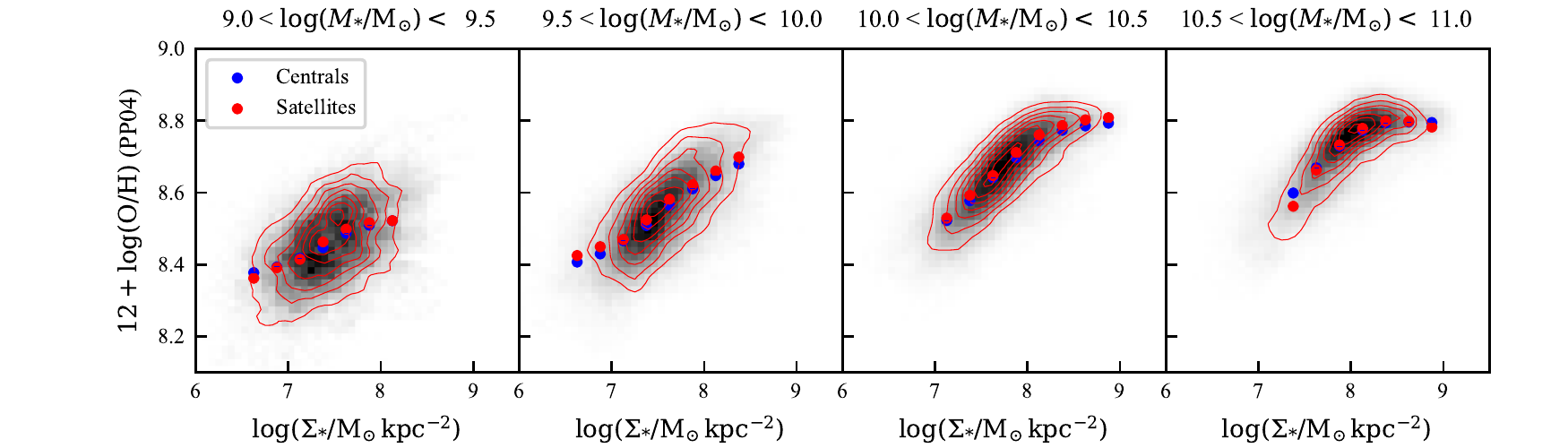}
\includegraphics{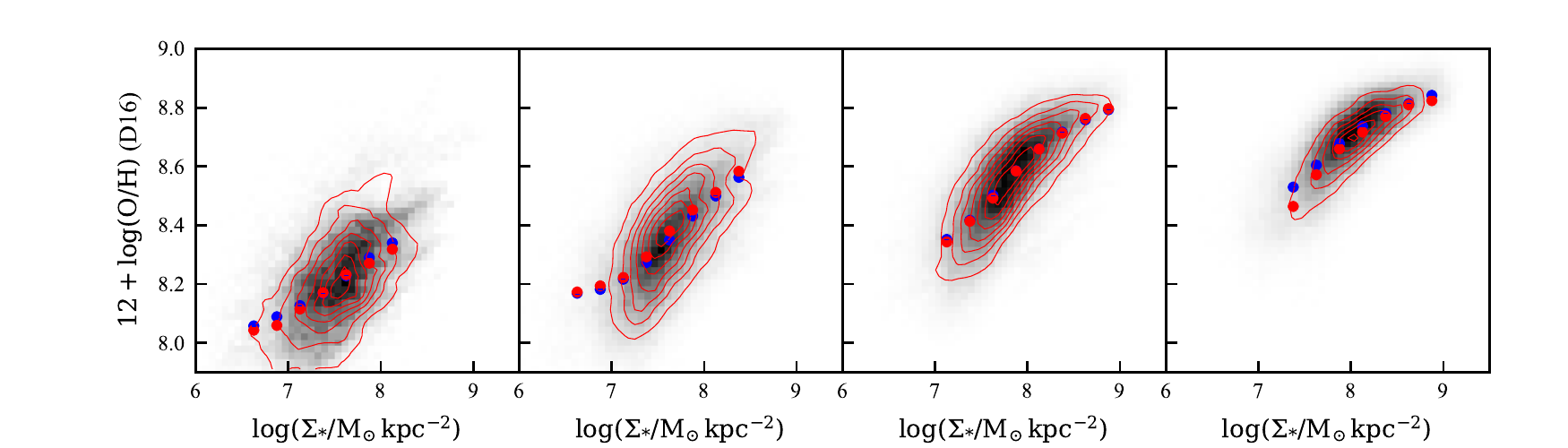}
\caption{The relationship between local stellar mass surface density and metallicity in bins of total stellar mass for satellite and central galaxies. The greyscale background represents the density of data points for central galaxies, while the red contours represent the distribution of data points from satellite galaxies. For clarity, the distribution for satellite galaxies was smoothed to make the contours less subject to noise. Blue points represent the median values of metallicity in the central galaxies at a fixed $\Sigma_{*}$, and the red points are the medians for satellite galaxies. These are only calculated where there are sufficient data. We include bootstrapped standard errors of the median, but these uncertainties are often smaller than the data points. In the top row we show the results for the \cite{Pettini2004} O3N2 indicator, while in the bottom row we show the result for the \cite{Dopita2016} N2S2H$\alpha$ indicator. The metallicity of satellite galaxies at fixed stellar mass surface density is slightly higher $(\sim 0.01 \, \mathrm{dex})$ than for central galaxies.}\label{sig_mass_met}
\end{figure*}

\subsubsection{Metallicity and Stellar Density}
As more stars form and the stellar surface density ($\Sigma_{*}$) increases, the amount of enrichment of the ISM also increases. We explore this relation in Figure \ref{sig_mass_met}, where we show the relationship between $\Sigma_{*}$ and metallicity for satellite and central galaxies. \cite{Hwang2019} showed that in addition to the relationship between the local $\Sigma_{*}$ and $12+\log(\mathrm{O/H})$, there is a secondary dependence on the total stellar mass. For this reason we have split our analysis into $0.5 \, \mathrm{dex}$-wide bins of integrated $M_{*}$. Using both the O3N2 and N2S2H$\alpha$, there is a small ($\sim 0.01 \, \mathrm{dex}$) difference between the metallicity at fixed $\Sigma_{*}$ between satellites and centrals, particularly in the $9.5 < \log(\mathrm{M_{*}/M_{\odot}})<10$ interval. While the formal uncertainties on the medians indicate that these differences are statistically significant, they are a factor of ten smaller than the standard deviations of the metallicity distributions.  We note that for systems of low stellar mass, a satellite galaxy may occupy a group with a wide range of possible halo masses, corresponding to very different environments.

\begin{figure}
\includegraphics{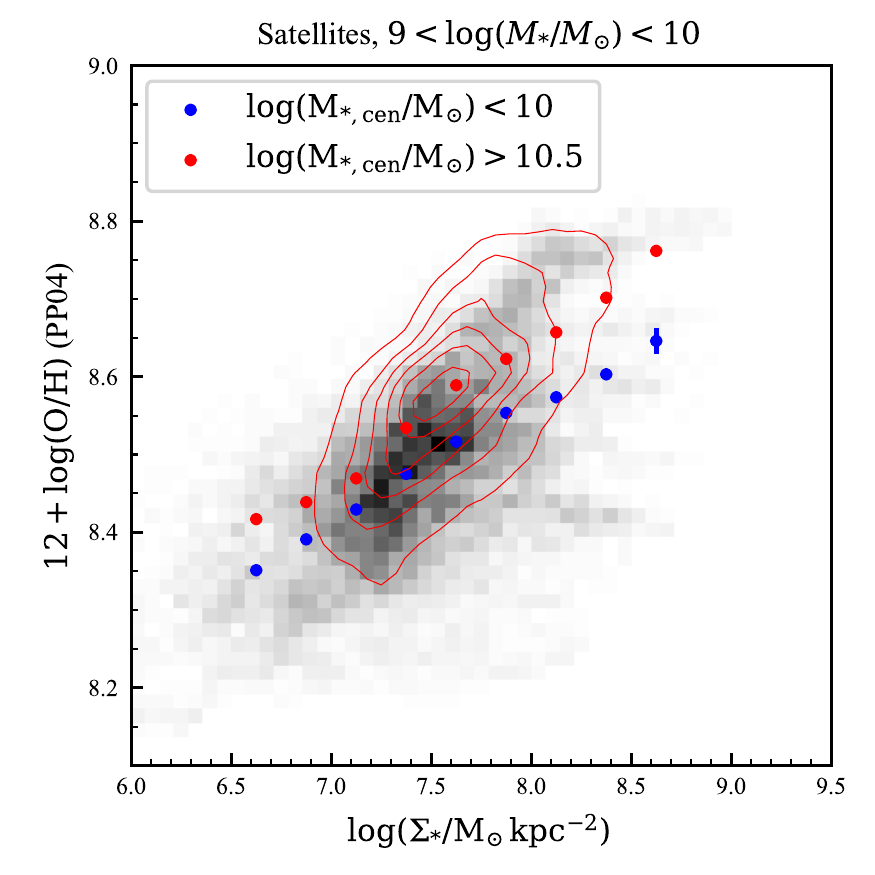}
\includegraphics{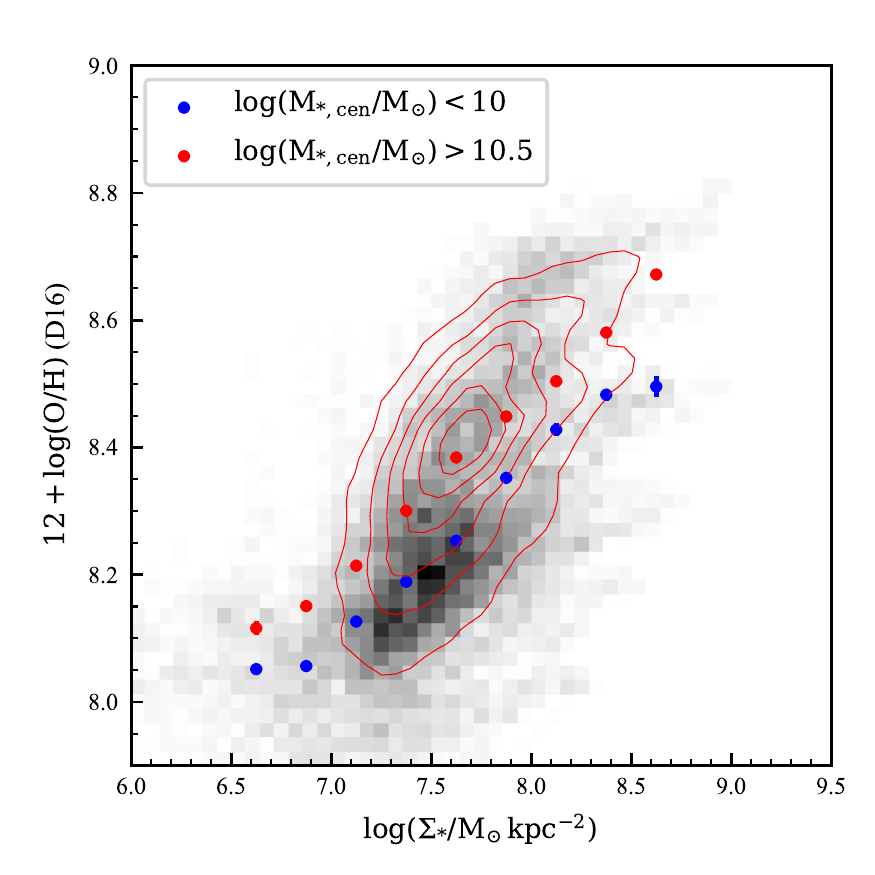}
\caption{The local $\Sigma_{*}-\mathrm{O/H}$ relation for satellite galaxies with $9<\log(M_{*}/\mathrm{M_{\odot}})<10$ split by the mass of the galaxy which is central to their halo. In the upper panel, we show the results for the PP04 indicator, and in the lower panel we show the results for the D16 indicator. Blue points show the median metallicity at a given $\Sigma_{*}$ for satellites of low-mass centrals ($\log(M_{*}/\mathrm{M_{\odot}})<10$). These points trace the median of the grey-shaded distribution. Red points are the median metallicity as a function of $\Sigma_{*}$ for satellites of high-mass centrals, shown by the red contours. The oxygen abundance is systematically higher for satellites of more massive centrals. }\label{sig_mass_met_cen_mass}
\end{figure}

Given that the largest differences in the metallicity between satellites and centrals occurs at the lowest stellar masses, we show the impact of varying the stellar mass of the central in Figure \ref{sig_mass_met_cen_mass}. Choosing satellite galaxies between $9<\log(\mathrm{M_{*}/M_{\odot}})<10$, we find a large systematic offset in metallicity for satellites of more massive central galaxies, corresponding to more massive group halos. Satellite galaxies associated with centrals more massive than $\log(M_{*}/\mathrm{M_{\odot}}) = 10.5$ have metallicities that are on average $0.08 \pm 0.009 \, \mathrm{dex}$ higher than those galaxies which are satellites of centrals with $\log(M_{*}/\mathrm{M_{\odot}}) < 10$. To eliminate the possibility that a different distribution of total stellar masses  for the targeted galaxies within the central stellar mass bin is responsible for the discrepancy, we perform a two-sample Kolmogorov-Smirnov test \citep{Smirnov1939}. This test returns a statistic of $D=0.17$ with $p=0.71$, indicating no statistically significant difference in the total stellar masses.

\subsubsection{Metallicity and gas fraction}
\begin{figure*}
\includegraphics{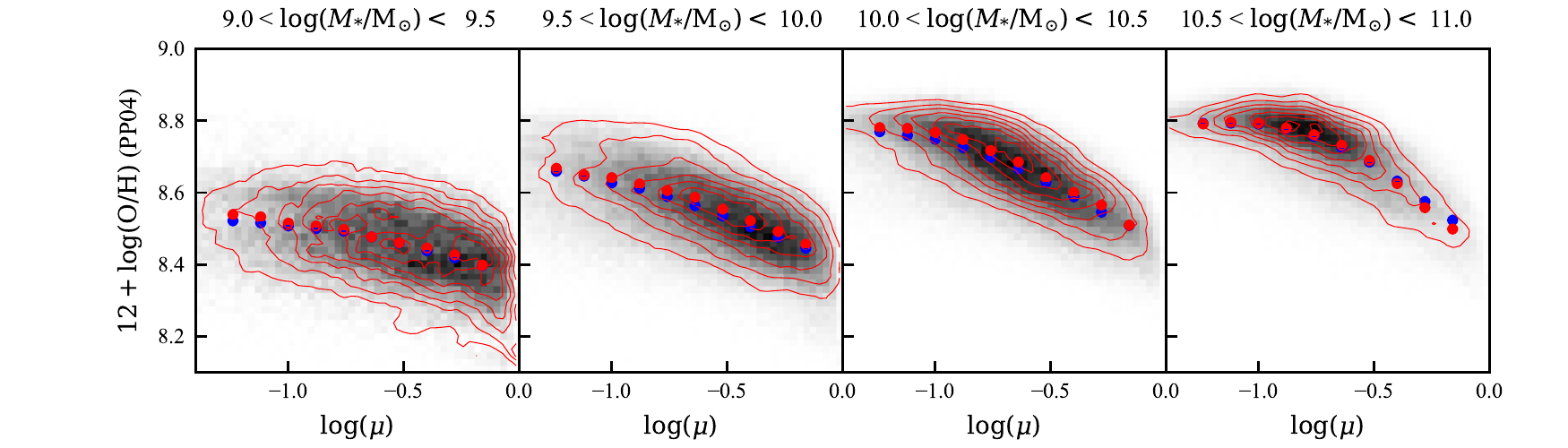}\\
\includegraphics{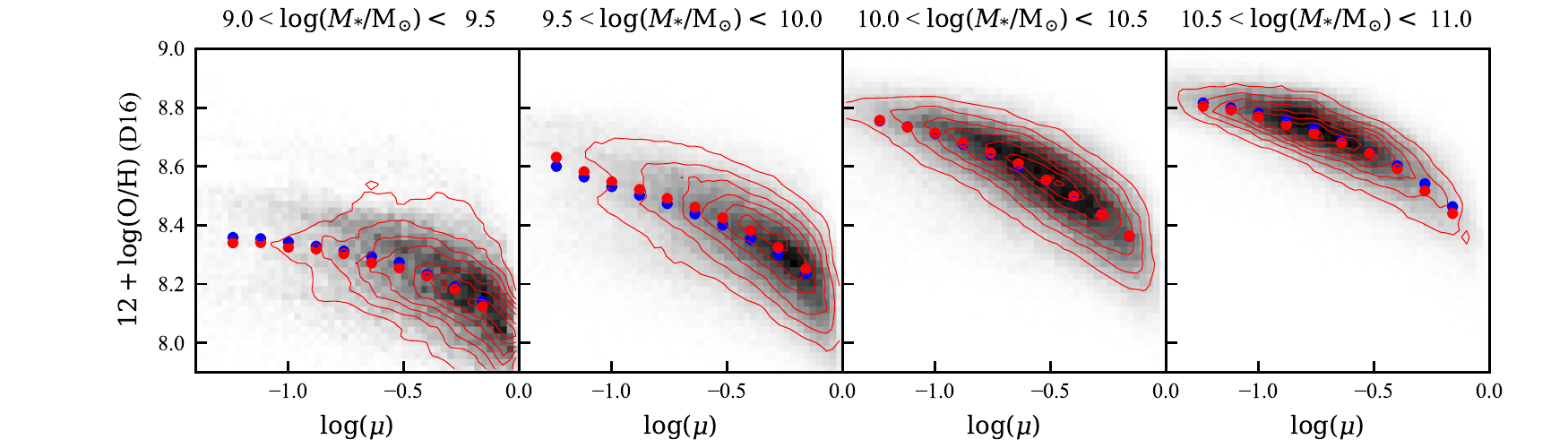}
\caption{The dependence of $12+\log(\mathrm{O/H})$ on the local gas fraction, $\mu$ in different bins of stellar mass. The contours, greyscale and coloured points are the same as in Figure \ref{sig_mass_met}. The difference in metallicity at fixed $\mu$ between satellites and centrals is largest in the range $9.5 < \log(M_{*}/\mathrm{M_{\odot}}) < 10$, where it reaches $\sim 0.015 \, \mathrm{dex}$.  }\label{mu_met}
\end{figure*}

Models predict \citep[e.g.][]{Lilly2013}, and observations show \citep{Mannucci2010, Moran2012}, that if low-metallicity gas is accreted onto the galaxy and the local gas fraction rises, then the metal content is diluted and the total metallicity of the gas will decrease. This relationship is investigated in Figure \ref{mu_met}, where we show the gas-phase metallicity as a function of the local gas fraction, $\mu$ in intervals of total stellar mass. In narrow bins of stellar mass, we find a tight correlation between the local gas fraction and the metallicity of the ISM. Once again, there is a small difference in the metallicities between satellites and centrals, with the difference being largest in galaxies between $10^{9.5}$ and $10^{10} \, \mathrm{M_{\odot}}$.

\begin{figure}
\includegraphics{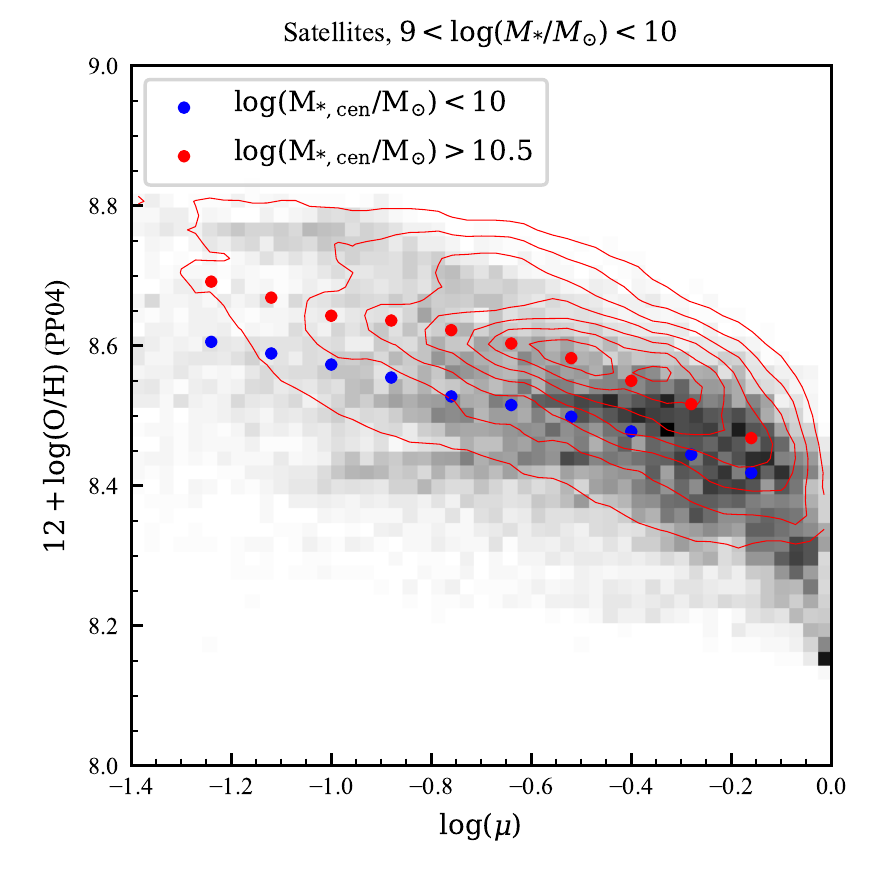}\\
\includegraphics{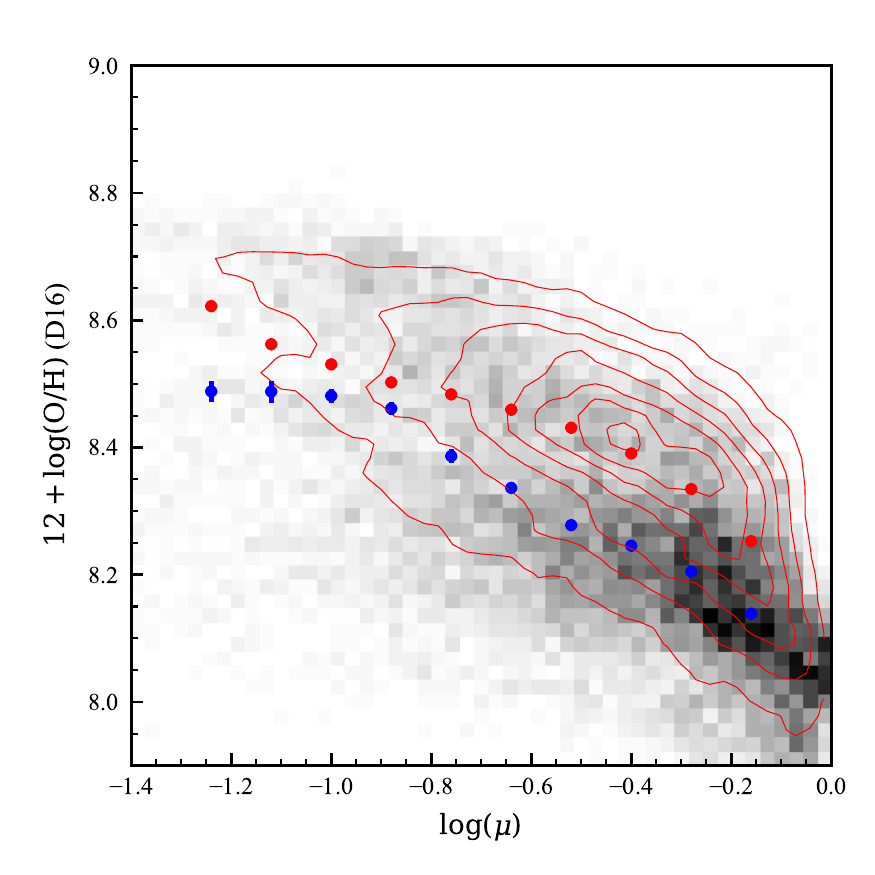}
\caption{The relationship between $\mu$ and O/H for satellites of low-mass centrals (grey background with blue points indicating the median) and high-mass (red contours with red points indicating the medians) for PP04 O3N2 (top) and D16 N2S2H$\alpha$ (bottom). At fixed gas fraction, the median metallicity is $\sim 0.1 \, \mathrm{dex}$ higher for the satellites of massive centrals.}\label{mu_met_cen_mass}
\end{figure}

Focussing again on the lower-mass satellite galaxies in our sample, we see in Figure \ref{mu_met_cen_mass} that the satellites of massive galaxies are more enriched at fixed $\mu$ than the satellites of less massive centrals. Comparing the contours of spaxels in the $\mu$-metallicity plane, we see that on average, the satellites of more massive centrals have a lower inferred gas fraction. Nevertheless, at fixed $\mu$ the offset in O/H remains.

\section{Discussion}\label{Discussion}
\subsection{The impact of environment on local scaling relations}
We have shown that the metallicity versus stellar mass, local escape velocity and gas fraction local scaling relations vary with the environment that galaxies inhabit. In this study we utilised the mass of the largest galaxy in the group as our estimate of environment. This quantity is correlated with the total mass of the halo \citep{Behroozi2010}, though does not suffer from the large uncertainties involved in estimating halo dynamical masses from spectroscopy \citep[see][for an excellent discussion of this point]{Robotham2011}. For satellite galaxies, the magnitude of the difference in metallicity appears to be a function of the stellar mass of the galaxy that is central to the group. In Figures \ref{sig_mass_met_cen_mass} and \ref{mu_met_cen_mass} we showed that the satellites of central galaxies more massive than $\log{\mathrm{M_{*}}/M_{\odot}}>10.5$ have local metallicities that are enhanced by $\sim 0.1 \, \mathrm{dex}$ over similar galaxies which are satellites of less massive ($\log{\mathrm{M_{*}}/M_{\odot}}<10$) centrals. This enhancement appears to be independent of the local gas fraction and escape velocity.

\subsection{Accounting for outflows with the gas-regulator model}
While differences in the metallicity of satellite galaxies at fixed $\Sigma_{*}$ and $\mu$ may be suggestive of some intrinsic difference between the chemical evolution of satellites in different mass halos, these simple scaling relations taken individually are unable to account for all factors that may influence the oxygen abundance. Neither of these scaling relations explicitly accounts for the loss of metals and corresponding reduction in oxygen abundance through outflows. To control for all of these factors at once, we fit the gas regulator model of \cite{Lilly2013} to the data. 

The gas regulator model for galaxy evolution makes the simple assumption that a galaxy's current star formation and rate and metallicity are largely determined by the present day gas fraction. While it was originally devised to apply to entire galaxies as a whole, some authors have recently showed that it can be applied to galaxies locally on $\sim \mathrm{kpc}$ scales \citep{Carton2015, BarreraBallesteros2018}. In their derivation of this model, \cite{Lilly2013} showed that the metal content of galaxies will reach an equilibrium on timescales shorter than the time it takes for their total gas content to be depleted. At equilibrium, the metallicity is

\begin{equation}\label{gas_regulator}
Z_{eq} = Z_{0} + \frac{y}{1 + r_{gas} + (1-R)^{-1} \left(\lambda + \epsilon^{-1}\frac{d \ln(r_{gas})}{dt} \right)},
\end{equation}
where $r_{gas}$ is the ratio of gas to stellar mass, $R$ is the fraction of gas returned from stars to the ISM by stellar evolution, and $\epsilon$ is the star formation efficiency. While this equation contains several unknown quantities, we can fix these to sensible values based on previous estimates from the literature. We adopt a value of $R = 0.4$, which is consistent with the predictions of stellar population synthesis models \citep{Bruzual2003}, and is in line with the assumptions underlying previous work on this topic \citep{Lilly2013,Carton2015,BarreraBallesteros2018}. Further, based on fitting the mass-metallicity relation for SDSS galaxies, \cite{Lilly2013} were able to constrain the product $ \epsilon^{-1}\frac{d\ln( r_{gas})}{dt} = -0.25$. The nucleosynthetic yield, $y$, is also not well known. The yield per stellar generation (and gas return fraction, $R$) is dependent on the stellar initial mass function, which some suggest may not be universal \citep[e.g.][]{Gunawardhana2011,Parikh2018}. \cite{Finlator2008} estimate the yield to be in the range $0.008 \leq y \leq 0.023$, but we assume the value near the middle of this range of $0.014$. This is the value calculated by \cite{BarreraBallesteros2018} based on both theoretical modeling using { \tt STARBURST99} \citep{Leitherer2014} and by closed-box modeling of galaxy cluster data \citep{Renzini2014}. We assume that this value is constant and valid throughout our entire sample. For a rigorous discussion of the impact of variations of the assumed yield on the calibration and interpretation of metallicities, see \cite{Vincenzo2016}.

In the gas regulator model, the outflows are described by the mass-loading factor, $\lambda$, which is the ratio of the star formation rate to the rate of mass loss due to stellar feedback and winds. We parametrized the mass-loading factor similar to the formulation of \cite{Peeples2011}, and assuming the metallicity of outflows is the same as the metallicity in the ISM of the galaxy

\begin{equation}\label{lambda}
\lambda = \left( \frac{v_{0}}{v_{esc}(r)}\right)^{\alpha}.
\end{equation}

\cite{Peeples2011} suggest either $\alpha=1$ or $2$, however we note that the they parametrise $\lambda$ in terms of the virial velocity of galaxy halos. The relationship between the virial velocity and the local escape velocity is complicated, and so to account for this we allow $\alpha$ to vary freely. \cite{BarreraBallesteros2018} also include an additive constant in their parametrisation of $\lambda$, which imposes a minimum level of outflows from even the deepest potential well. For a reasonable choice of yield, we find that this has the effect of limiting the maximum metallicity that a gas-regulated system can achieve.

\subsubsection{Fitting the \cite{Pettini2004} metallicity}
In order to constrain the values of $v_{0}$ and $\alpha$ for our model, we fit equation \ref{gas_regulator} to the metallicity, gas fraction and local escape velocities inferred from the MaNGA spaxel data for all galaxies in our sample. We find $v_{0}=368 \, \mathrm{km \, s^{-1}}$, and $\alpha = 0.52$. In their work on deriving the $v_{esc}$-dependence of $\lambda$, \cite{BarreraBallesteros2018} noted that the gas regulator model does not necessarily provide a good fit to the data, though it was preferred to the leaky-box model of \cite{Zhu2017} on the grounds that it provided more realistic estimates of $\lambda$. In our fits of the gas-regulator model to the MaNGA data on kpc scales, we find smaller residuals at low gas fraction. This is a direct result of our choice not to include an additive constant in our parametrisation of $\lambda$. In the analysis that follows we fix the dependence of $\lambda$ on the escape velocity, reducing this problem to fitting only one variable, the metallicity of accreted gas, $Z_{0}$.

We perform a least-squares fit of the gas regulator model to the O3N2-based gas-phase metallicities, fitting for $Z_{0}$ in the sub-populations where the largest difference in metallicity is seen. This is for satellite galaxies with stellar masses in the range $9<\log (M_{*}/\mathrm{M_{\odot}})<10$, split based on the stellar mass of the corresponding central galaxy. For satellites of low-mass centrals ($\log(M_{*}/\mathrm{M_{\odot}})<10$), the metallicity of the gas precipitating onto their discs inferred from the modeling is $Z_{0} = (4.68 \pm 0.11) \times 10^{-4}$, corresponding to $12+\log(\mathrm{O/H})=7 .46 \pm 0.01$. For the gas being accreted onto the $9<\log(M_{*}/\mathrm{M_{\odot}})<10$ satellites of high-mass central galaxies, we derive a metallicity of $Z_{0}= (1.17 \pm 0.001) \times 10^{-3}$, or $12+\log(\mathrm{O/H})=7.87 \pm 0.003$.

\subsubsection{Fitting the \cite{Dopita2016} metallicity}
The metallicities derived from the \cite{Dopita2016} N2S2H$\alpha$ calibration have a different absolute abundance scaling and cover larger range for the same set of spectra. This can be seen by comparing the $y$-axes in Figure \ref{sig_mass_met}. Using the same values of the yield and gas return fraction as were used to fit the O3N2 metallicities, the data favours a negative value of $Z_{0}$, which is unphysical. With this indicator, the values of $\lambda$ allowed by this parametrisation that also gives an appropriate shape to the distribution of modeled data in the $\mu -Z $ plane, are too large. Using the $\lambda$ parametrisation of \cite{BarreraBallesteros2018}, $\lambda=\left(v_{0}/v_{esc} \right)^{\alpha} + \lambda_{0}$, we find $Z_{0} = (3.0 \pm 0.12 )\times 10^{-4}$ or $12+\log(\mathrm{O/H})=7.27 \pm 0.01$ for the satellites of low-mass centrals. For the satellites of high-mass centrals we find  $Z_{0} = (8.6 \pm 0.1 )\times 10^{-4}$ or $12+\log(\mathrm{O/H})=7.86 \pm 0.003$.

We note that the absolute value for these inferred quantities is correlated with a number of unconstrained parameters including the nucleosynthetic yield of oxygen, $y$, the gas return fraction from stars, $R$, and the precise form of the mass loading factor, $\lambda$. Nevertheless, we argue that the assumption that these parameters do not vary between star-forming galaxies in a relatively narrow mass range is reasonable, and that the choice to fix them for this comparison is justified. With this limitation it is not possible to derive an absolute abundance for the accreted gas, but the existence of a difference is robust. As was shown in Figures \ref{sig_mass_met_cen_mass} and \ref{mu_met_cen_mass}, the difference in local metallicity scaling relations is significant between these two galaxy subpopulations. The gas regulator model provides an interpretive framework to describe these differences in terms of the variation of the metallicity of the intergalactic medium in different environments, while controlling for small differences in the estimated local gas fraction and escape velocity.

\subsection{Can starvation explain our results?}\label{Starvation}
Our results are analogous to those seen by previous studies of the environmental dependence of the global mass-metallicity relation \citep[e.g.][]{Cooper2008,Pasquali2012,Peng2014,Wu2017}. While different studies have found qualitatively similar results, there is considerable disagreement in the interpretation. \cite{Peng2014}, argue that the primary driver of this trend must be the elevation of the metallicity of gas being accreted onto galaxies in dense environments. This argument hinges on their observation that the distribution of star-formation rates in their sample is independent of environment.
This conclusion contrasts starkly with the interpretation of \cite{Wu2017}, who suggested that the environmental variation of the mass-metallicity relation can be explained by the reduction in the gas-fractions of galaxies with the local galaxy overdensity. 

\begin{figure}
\includegraphics{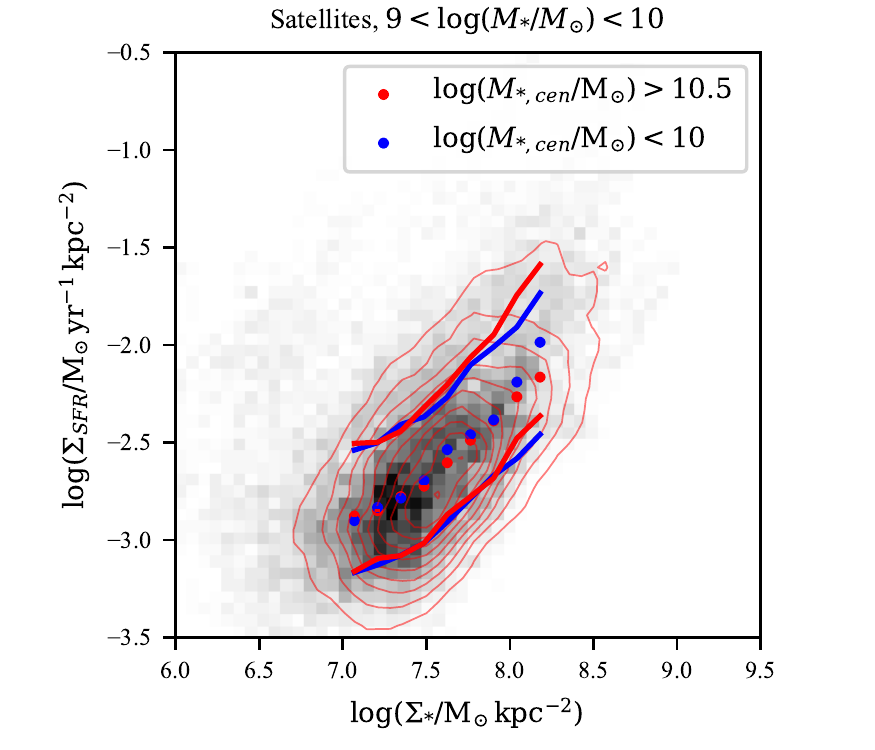}
\caption{The star formation rate surface density as a function of stellar mass surface density for galaxies with $9<\log(M_{*}/\mathrm{M_{\odot}})<10$. The greyscale shows the distribution of measurements from satellites of low-mass centrals ($\log(M_{*,cen}/\mathrm{M_{\odot}})<10$), with the median of this distribution shown by blue points, and the $16$th and $84$th percentiles shown by blue lines. The red contours indicate the $\Sigma_{*} - \Sigma_{SFR}$ distribution for satellites of massive galaxies ($\log(M_{*,cen}/\mathrm{M_{\odot}})>10.5$), with the red points showing the median and the red lines marking the $16$th and $84$th percentile of the distribution. There is very little difference in the two distributions for the vast majority of spaxels in the two samples.}\label{sig_M_sig_SFR_cen_mass}
\end{figure}

Starvation, whereby the accretion of gas onto galaxies is curtailed and the gas reservoir is not replenished following star formation \citep{Larson1980}, will have the effect of increasing the gas-phase metallicity of a galaxy or a region of the galaxy. This is a natural consequence of maintaining a constant metal yield from stellar evolution, while reducing the replenishment of the reservoir with relatively low-metallicity gas. Within the framework of gas-regulated galaxy evolution, this implies an anti-correllaton between the gas surface density or star formation rate surface density and the metallicity in the gas. Starvation has been suggested as a key component for determining the star-forming properties of galaxies today \citep{Peng2015,Trussler2018}, with environment appearing to play a role in instigating this process \citep{vonderlinden2010, Davies2016}. 

While we do infer gas fractions that are, on average, lower for galaxies that are in more extreme environments (for example low-mass satellites of high-mass galaxies), we find that at fixed gas fraction the metallicity is higher for satellites relative to centrals, even in spaxels with high $\mu$. The differing distributions of $\log(\mu)$ evident in Figure \ref{mu_met_cen_mass} are largely driven by the differences in the distributions of $\Sigma_{*}$. Although the distributions of total $M_{*}$ between the two subsamples used are not significantly different, the distributions of local stellar mass surface densities are. In Figure \ref{sig_M_sig_SFR_cen_mass} we show the joint distributions for $\Sigma_{*}$ and $\Sigma_{SFR}$ for low-mass satellites, split by the stellar mass of the central galaxy. At fixed $\Sigma_{*}$, the difference between the means of the $\Sigma_{SFR}$ is smaller than $0.03$ dex, except above $\log(\Sigma_{*})=8.1$, but this range accounts for only $\sim 10 \% $ of the data and therefore has a minimal impact on our model fitting. 

It is possible that the distribution of star formation has also changed. \cite{Schaefer2019} showed that in dense environments, the outer parts of a galaxy can be quenched, leaving star formation in the inner regions unaffected. This transformation was nevertheless accompanied by a reduction in the total specific star formation rate (sSFR). To test for this, we perform a KS-test on the integrated specific star formation rates of the two subsamples. This yields  $KS = 0.17$ with $p=0.58$. Furthermore, the median of the sSFR for the satellites of high mass galaxies is $-10.12 \pm 0.07 \, \mathrm{yr^{-1}}$ and the median for the satellites of more massive galaxies is $-10.21 \pm 0.04 \, \mathrm{Gyr^{-1}}$, where the error on the median has been estimated using a bootstrap resampling. The difference of medians is within the error margin. 

The similar distributions of $\Sigma_{SFR}$ and  sSFR disfavour the interpretation that the changing gas fraction due to starvation is responsible to the environmental differences in metallicity on kpc scales within our sample. This is not to say that starvation does not occur in dense environments; our sample selection simply favours the most star-forming galaxies, which are unlikely to have had their star formation rates reduced by environmental effects yet. Environmental differences in metallicity may occur in satellite galaxies before the onset of environment quenching.

\subsection{Comparison to simulations}
Numerical simulations of galaxy evolution are beginning to show that the gas being accreted onto galaxies cannot be assumed to be pristine in all environments \citep{Oppenheimer2010,Gupta2018}. In the simulations, the origin of accreted gas is observed to be highly dependent on redshift, with cosmological accretion of low-metallicity gas dominating at high redshift. However, as time progresses, feedback from star formation and AGN activity expel gas from the interstellar media of galaxies, which enriches their local environment with material that subsequently falls onto their neighbours. In the FIRE simulations, \cite{AnglesAlcazar2017} found that the exchange of gas between galaxies that is facilitated by galactic winds dominates the accretion budget by $z=0$. 

\cite{Gupta2018} explored this effect using data from the IllustrisTNG simulations. They showed that the enrichment of the intergalactic medium and the associated accretion onto galaxies is dependent on both the halo mass and whether a galaxy is infalling into its host halo or whether it has been a satellite for some time. At $z<0.5$, they find that the metallicity of gas being accreted onto galaxies with $9<\log(M_{*}/M_{\odot})<10$ that are infalling into clusters is approximately $0.35 \, Z_{\odot}$, which is $1.5$ - $2$ times more metal rich than for similar galaxies in the field. This is consistent with the metallicity difference that we have inferred between the satellites of low-mass and high-mass centrals, though the absolute abundances differ. We again note that the value of $Z_{0}$ returned when the model represented by Equation \ref{gas_regulator} is fitted to the data is sensitive to the precise values of the yield, $y$, and the gas return fraction, $R$, which are not well constrained by observations. The choice of these values will change the estimate of $Z_{0}$, but the relative difference between subsamples will not be greatly affected.

\subsection{Other studies of the environmental dependence of metallicity}
The impact of environment on galaxy evolution, in particular star formation and metallicity, is subtle. For this reason there have been very few observational works that measure the impact of environment on the spatial patterns of chemical abundances in galaxies. It has only been recently that large enough samples of integral field spectroscopic data have become available to adequately measure these effects. 

In a recent study, Lian et al. (\emph{in prep.}) used MaNGA data to study the metallicity gradients of galaxies as a function of the local environmental overdensity. They find that the metallicity gradients in low-mass satellite galaxies are shallower in dense environments, with a higher metallicity in their outer parts than similar galaxies in the field. They also find that the star formation rate gradients in galaxies in dense environments are steeper and conclude that the most likely explanation for these observations is a variation in the gas accretion timescale in different environments. Superficially this would seem to contradict our results, but we argue that this apparent disagreement can be resolved by noting the differences between the samples of galaxies considered. Lian et al. place less stringent constraints on the number of star-forming spaxels than we do, meaning that their galaxies have lower specific star formation rates on average. They therefore study galaxies that are likely to have inhabited their host halos for a longer period of time, and are more affected by environment quenching processes. This point is made in Section \ref{Starvation}, where we rule out starvation as the primary driver of the environmental effects discussed in this work.  Additionally, we note that Lian et al. placed no constraints on the inclination of galaxies in their sample to the line of sight. This may explain the differences in the metallicity gradients to those reported in our work.

\section{Conclusions}
We have estimated local metallicities, gas fractions, escape velocities and star formation rate surface densities for a sample of nearly face-on star-forming galaxies observed by MaNGA. In this sample we have explored the impact of the environment on local scaling relations between these estimated quantities, with a particular focus on satellite galaxies. 
We find
\begin{itemize}
\item{At fixed stellar mass, we find a small but global offset of $0.025 \, \mathrm{dex}$ (for O3N2) or $0.05 \, \mathrm{dex}$ (for N2S2H$\alpha$) in the metallicities of galaxies between satellites and centrals. For our sample we find little evidence for changes in the metallicity gradient between satellites and centrals.}
\item{The disparity between the metallicity of satellites and centrals is also evident in the O/H -- $\Sigma_{*}$ and O/H -- $\mu$ local scaling relations. We find the greatest offset when we split our satellite sample by the stellar mass of the galaxy that is central to the respective halo. For satellite galaxies in the range $9<\log(M_{*}/M_{\odot})<10$, the local scaling relations are $\sim 0.1 \, \mathrm{dex}$ more oxygen rich for satellites of hosts more massive than $10^{10.5} \, M_{\odot}$ than for hosts less massive than $10^{10} \, M_{\odot}$.}
\item{The offset in metallicity for satellite galaxies is found to exist between different environments at constant stellar mass surface density, gas mass fraction and star formation rate surface density. From these we conclude that the observed differences cannot be explained by gas starvation occurring in satellites around more massive centrals. Interestingly, the impact of environment on the chemical enrichment of galaxies appears to precede the onset of the quenching of star formation in their disks. }
\item{Measured on kpc-scales, local metallicities and gas fractions are found to be quantitatively consistent with the gas regulator model of \cite{Lilly2013}. We assume that the mass loading factor describing outflows in galaxies is a function of the local escape velocity. Within the framework of the gas regulator model the only explanation for the elevated metallicity is an increase in $Z_{0}$ for satellites of high-mass galaxies. We estimate that the oxygen abundance in the inflowing gas changes from $12+\log(\mathrm{O/H}) = 7.54 \pm 0.01$ to $7.86 \pm 0.003$ using the \cite{Pettini2004} O3N2 indicator, or $12+\log(\mathrm{O/H}) = 7.27 \pm 0.01 $ to $7.86 \pm 0.003 $ for N2S2H$\alpha$.}

\end{itemize}

Given these conclusions, we interpret the enhanced metallicity of the satellites of more massive centrals to be evidence for the exchange of enriched gas between galaxies. In this picture, which has been motivated by both observations \citep{Peng2014} and simulations \citep{Oppenheimer2010,AnglesAlcazar2017,Gupta2018}, feedback driven winds expel metal-rich gas from a massive star-forming central which is subsequently accreted onto nearby satellites. While our estimates for the metallicity of gas accreted onto satellite galaxies are a factor of $\sim 3$ lower than the predictions of \cite{Gupta2018}, we note that the values returned by our modeling are subject to inherent uncertainties in the nucleosynthetic yield, the gas return fraction from stellar evolution, and the absolute abundance scaling of the strong-line metallicity diagnostics. Notwithstanding these systematic effects, the inferred differential in the metallicity of gas accreted onto satellites in different environments is qualitatively in good agreement with the simulations.

The impact of environment on the gas phase metallicity distribution of galaxies is likely to be complicated and multifaceted. In addition to the accretion of enriched gas in dense environments that we have studied here, tidal interactions, mergers and ram pressure can influence the distribution of metals in a galaxy. A complete understanding of the effect of environment on gas phase metallicities must take these other processes into account. A more comprehensive analysis of the detailed environmental dependence of the chemical properties of galaxies will be made possible when the full MaNGA sample becomes available, or with future integral field spectroscopic surveys such as HECTOR \citep{Bryant2016}.

\acknowledgements 
We would like to thank the anonymous referee, whose constructive comments were extremely helpful in clarifying some aspects of this paper. 
ALS, ZJP and CT acknowledge NSF CAREER Award AST-1554877. AJ acknowledges NSF Award 1616547. RM acknowledges ERC Advanced Grant 695671 "QUENCH" and support from the the Science and Technology Facilities Council (STFC).

This research made use of \texttt{Astropy}, a community-developed core \texttt{python} package for astronomy \citep{Astropy2013,Astropy2018}; \texttt{matplotlib} \citep{Matplotlib}, an open-source \texttt{python} plotting library; and \texttt{LMFIT} \citep{LMFIT}, an interface for non-linear optimization in \texttt{python}.
Funding for the Sloan Digital Sky Survey IV has been provided by the Alfred P. Sloan Foundation, the U.S. Department of Energy Office of Science, and the Participating Institutions. SDSS acknowledges support and resources from the Center for High-Performance Computing at the University of Utah. The SDSS web site is www.sdss.org. SDSS is managed by the Astrophysical Research Consortium for the Participating Institutions of the SDSS Collaboration including the Brazilian Participation Group, the Carnegie Institution for Science, Carnegie Mellon University, the Chilean Participation Group, the French Participation Group, Harvard-Smithsonian Center for Astrophysics, Instituto de Astrof\'{i}sica de Canarias, The Johns Hopkins University, Kavli Institute for the Physics and Mathematics of the Universe (IPMU) / University of Tokyo, the Korean Participation Group, Lawrence Berkeley National Laboratory, Leibniz Institut f\"{u}r Astrophysik Potsdam (AIP), Max-Planck-Institut f\"{u}r Astronomie (MPIA Heidelberg), Max-Planck-Institut f\"{u}r Astrophysik (MPA Garching), Max-Planck-Institut f\"{u}r Extraterrestrische Physik (MPE), National Astronomical Observatories of China, New Mexico State University, New York University, University of Notre Dame, Observat\'{o}rio Nacional / MCTI, The Ohio State University, Pennsylvania State University, Shanghai Astronomical Observatory, United Kingdom Participation Group, Universidad Nacional Aut\'{o}noma de M\'{e}xico, University of Arizona, University of Colorado Boulder, University of Oxford, University of Portsmouth, University of Utah, University of Virginia, University of Washington, University of Wisconsin, Vanderbilt University, and Yale University.

\bibliography{bibliography}

\label{lastpage}

\end{document}